\documentclass{osa-article}

\journal{oe}
\articletype{Research Article}
\begin{document}

\title{Light scattering through the graphene oxide liquid crystal in a micro-channel}

\author{M. Arshadi Pirlar,\authormark{1,2}  M. Rezaei Mirghaed,\authormark{1} Y. Honarmand,\authormark{1} S. M. S. Movahed,\authormark{1,2} and R. Karimzadeh\authormark{1,2,*}}

 \address{\authormark{1}Department of Physics, Shahid Beheshti University, Velenjak, Tehran 19839, Iran\\
 \authormark{2}Ibn-Sina Multidisciplinary Laboratory, Department of Physics, Shahid Beheshti University, Velenjak, Tehran 19839, Iran}
\email{\authormark{*}r$\_$karimzadeh@sbu.ac.ir}

\begin{abstract}
In this paper, we examine the light scattering by the flow of levitated flakes in a micro-channel to characterize the tunable functionality of the graphene oxide liquid crystal in the nematic phase. Light interaction with the mentioned material is decomposed to the scattered and transmitted parts and they can determine the orientation of the flakes. Our results demonstrate that, pumping the graphene oxide sample through the micro-channel leads to increase the amplitude of scattered light. The time averaged of scattered light intensity grows by increasing volume fraction. We also find that, the higher volume fraction, the sooner reaching to saturated normalized scattered intensity is. To get deep insight about our experimental results, we rely on the general theoretical properties of the light scattering cross-section incorporating the fluctuation of director vector and dielectric tensor. Our proposal is a promising approach to carry out  the mechanical-hydrodynamical approach for controlling the orientation of a typical liquid crystal.  
\end{abstract}
  
%{\bf Keywords}: Flow rate, Scattering, Anisotropic medium, Nematic Liquid Crystal, Graphene Oxide.

%\maketitle

\section{Introduction}\label{intro}
Liquid crystals have provided great opportunities based on corresponding optical, electrical and magnetic properties from fundamental and applied  science points of view  \cite{Khoo07, Blinov11}. Alignment of liquid crystal (LC) molecules through the so-called {\it director} direction, ${\bf n}$, producing a nematic phase. 
Two-dimensional (2D) nano-materials with dynamically tunable liquid crystalline properties have recently been appeared as a highly-promising class of novel functional materials. In order to produce a typical LC made of aqueous plate-like material, a proper 2D material particles are dispersed into a conventional liquid host with certain ranges of concentration \cite{Fu,Hogan17-1,Gao11,Zakri13,Hogan17}. 
Among the plate-like materials, graphene oxide (GO) has received most attention due to its unique properties such as availability, easy to make and its application in various area such as solar cells, circuit boards, display panels and biotechnology \cite{Randviir14}. Dispersions of GO in the water within a specific range of volume fractions indicates a nematic liquid crystal (NLC) phase known as GO-NLC \cite{Hogan17}.    

 Many different methods have been used to manipulate the orientation and ordering of graphene-based materials, including flow shear and also applying electric or magnetic fields \cite{Blinov11,Pleiner83,Marusii86,He15,Song16,Ahmad15,Ahmad16,Hong15,Nehring76,Ong84,Brochard72,Doorn75,Bao17}.  The controllability  by utilizing an electric or magnetic field is more feasible than mechanical-based methods such as flow shear approach proposing more flexibility and it is more suitable for device applications \cite{Nehring76,Ong84,Brochard72,Doorn75,Ahmad15,Ahmad16,Hong15}. The Frederiks transition and electro-optical characteristics have been examined by applying electric or magnetic field on LC \cite{Nehring76,Ong84,Brochard72,Doorn75}.  Also the amount of anisotropy in GO in response to an induced electric field  has been examined in \cite{Song16,Ahmad15,Ahmad16,Hong15}. Particularly, application of electric field on the GO-NLC leads to an undesired accumulation of GO at the cathode \cite{kim11}, subsequently alternative methods for rearranging the GO-NLC have been suggested \cite {Wang09,Lin15,Lin17,kim11}. It has been demonstrated that GO has almost weak response with respect to external magnetic field, consequently, to rearrange the corresponding directors, we need to high magnetic field causing to use alternative mechanism for doing alignment \cite {Wang09,Lin15,Lin17,kim11}.

Recently, Cheng \textit{et al.} demonstrated that GO-NLC can be used as a rewritable medium for reflective display without polarizing optics \cite{He15}. They have also shown that, the GO-NLC outer layer becomes reflective and bright, and they manually drawn patterns with dark appearance by applying a typical stick. Indeed, by imposing the stress on the top of sample by stick, the orientation of the GO flakes can change. It turns out that, they did not have any specific control over the amount of created stress in order to make a typical pattern.  Therefore, the notion of controllability of director orientation in a typical LC by using mechanical approaches has considerable challenge \cite{Pleiner83,Marusii86,He15,Sengupta13,Batista15}.  According to the hydrodynamical properties of NLC's considered in \cite{Khoo07, Blinov11, Marusii86, Leslie85, Pleiner83, Sengupta13}, one can combine physical behavior and boundary conditions to get deep insight in controllability of NLC materials.

A bundle of the light beam can be scattered by the fluctuations of LC molecules in a NLC \cite{Khoo07, Blinov11, deGennes77, Poggi77, Martinand72, Reznikov84, Marusii86, Leslie85, Pleiner83}. The dynamical properties governing on the light scattering \cite{Khoo07, Blinov11, deGennes77} and the effect of external fields on the mentioned phenomenon have been investigated \cite{Poggi77, Martinand72, Reznikov84, Marusii86, Leslie85, Pleiner83}. It has been depicted that, the non-equilibrium steady state of NLC has an impact on the light scattering cross-section \cite{Marusii86}. Also, the non-equilibrium conditions in the micro-channels can be controlled by the flow rate and boundary conditions. Therefore, we can manipulate intensity and the wavelength of the scattered light  by changing the flow rate in the channel. In this paper, we are motivated to propose a reliable and robust experimental setup to control  the optical anisotropy leading to change the light scattering cross-section based on GO-NLC in a micro-channel. Subsequently, our approach has the following advantages and novelties: proposing a new mechanical approach to achieve robust controllability mechanism in a simple experimental setup comparing to that of considering an external electric or magnetic field. Furthermore, we can increase the precision of controllability in a proper mechanical pipeline.   
This approach can be used in optical switching systems, smart windows, optical limiting and reflective display.

The rest of this paper is organized as follows: in section \ref{experiment1}, we will clarify our experimental setup and sample preparation. Section \ref{result1} will be devoted to the controllability of light scattering through the GO-NLC in a micro-channel. Summary and conclusion will be given in section \ref{summary1}

\section{Setup and material}\label{experiment1}
In this section, we clarify the  materials and experimental setup utilized for examining the light scattering through the GO-NLC injected in a typical micro-channel. We will also turn to the matter preparation and characterization for the rest of our experiment.

\subsection{Graphene oxide preparation and properties}
As mentioned in the introduction, we are interested in examining the effect of GO-NLC flow rate in a typical micro-channel on the bundle of light scattering. At first, we make GO by a modified Hummers method (prepared at $0.17$ and $1.00$mg mL$^{-1}$ concentrations in the water). The UV- visible, FTIR spectra, and the XRD pattern have been utilized to characterize GO. The size distribution of GO plates can be determined by using the DLS method.

\begin{figure}[t]
		\centering
		\includegraphics[width=9cm, height=5.5cm]{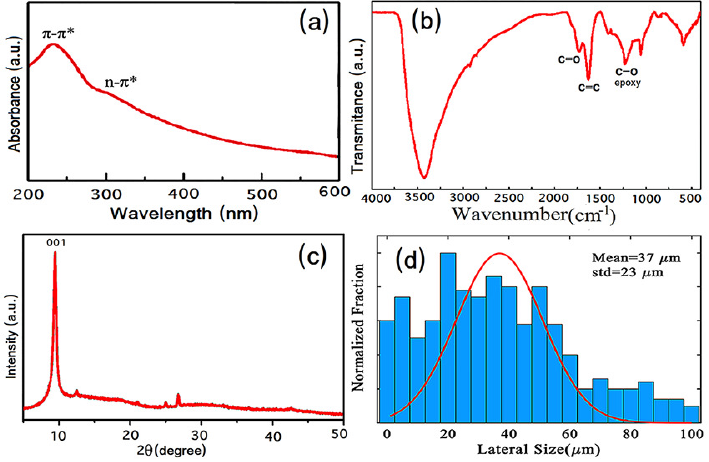}
		\caption{\label{fig:chrac}The characteristic spectra of the graphene oxide. a) uv-visible spectrum, b) the FTIR spectrum, c) the XRD pattern and d) the size distribution graph using the DLS method}
		
\end{figure}

Figure \ref{fig:chrac}(a) illustrates the UV absorption spectrum of GO indicating a shoulder peak at 300nm (due to $ n-\pi^*$
transitions of aromatic C-C bonds) and a maximum absorption peak at $232$nm (due to $\pi-\pi^*$
transition of the atomic C-C bonds \cite{Xu13}). The FTIR spectrum of GO shown in Fig. \ref{fig:chrac}(b) confirms
the absorption peaks of $1730$cm$^{-1}$ (carboxyl groups), $1061$cm$^{-1}$ (epoxy groups), $1385$cm$^{-1}$ (C-O
vibration), $1630$cm$^{-1}$ (C=C vibration), $740$cm$^{-1}$ (C=O vibration) and $3450$cm$^{-1}$ (O-H vibration)
\cite{Guo09}. The XRD pattern of GO is also shown in Fig. \ref{fig:chrac}(c). An interlayer distance of $0.77$nm
corresponds to GO can be obtained from the sharp diffraction peak observed at $2\theta=9.02^{\circ}$ \cite{Du10}.
Figure \ref{fig:chrac}(d) shows the size distribution of graphene sheets produced by using the DLS
technique. 
\begin{figure}[t]
		\centering
		\includegraphics[width=6cm, height=2.5cm]{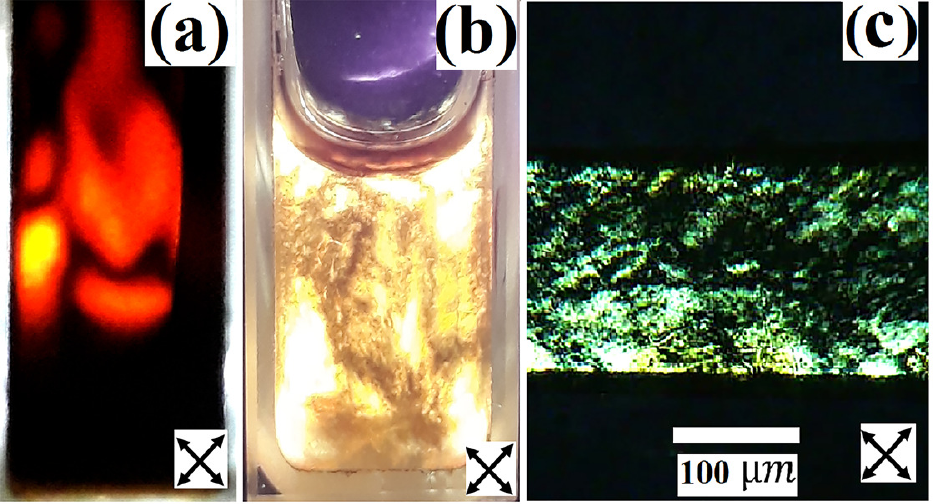}
		\caption{\label{fig:aniso}The passing light through the GO, which is under the influence of external force and placed between the crossed polaroid. The thickness of the GO sample is (a) 1 cm and (b) 0.1 cm. (c) GO within a micro-channel with a rectangular cross section of $300 \times 100 \mu$m. }
		
\end{figure}

We should also confirm that  our prepared GO-LC has the nematic phase. In other word, we have to
demonstrate that whether our produced material is a liquid crystal with an anisotropic property or not. By specifying the mean size of the GO's flakes and associated standard deviation, we determine the theoretical volume fractions for phase transitions from the isotropic phase to the biphase ($C_{IB} =0.03$Vol$\%$), and from the biphase to the nematic phase ($C_{BN} = 0.1$Vol$\%$) using Onsagers theory \cite{shen14}  which is an approximation due to the broadness of flakes size distribution.

 To check whether our prepared aqueous GO with the volume fraction of about $C=0.1$Vol$\%$ or more has the nematic phase, we pour the GO (at volume fraction of $C=0.17$Vol$\%$) into a $1\times 1$cm$^2$ square cell and placed it between crossed polarizers. If the suspended GO's flakes in a solvent are randomly oriented, which means isotropic phase, the exposure light does not feel birefringence, therefore, the light does not pass through the analyzer.
When we create a perturbation in a sample by an external factors, the directors are aligned locally and create an optical anisotropic in GO-NLC. The external factors can be considered as the external electric or magnetic fields applied on the sample, temperature gradient, mechanical stress such as pulling a stick, creating a flow in the sample, or creating a vibration to the cell containing GO-NLC sample. In this experiment, we created a tension in the GO-NLC dispersion by adding a drop of GO-NLC to the cell containing the sample (see supplementary information Visualization 1). Thus, the polarization of the light passed through the anisotropy GO-NLC is changed. Therefore, light can be detected after analyzer. In Figs. \ref{fig:aniso}(a) and \ref{fig:aniso}(b), we show the image of GO-NLC with a volume fraction of $C=0.17$Vol$\%$  between the crossed polarizers. It is worth noting that, the type of color at each point of sample observed after analyzer, depends on the variation of light polarization. 
The thickness and birefringence of sample cause the variation of polarization at different wavelengths \cite{Hong15,Lin15,kim11,Xu11}. 
 In our experimental setup, the mentioned variations are due to the induced optical anisotropy produced by applying the external factors. In this case, we change the thickness of our sample and set to $0.1$ cm. By comparing Fig. \ref{fig:aniso}(a) with Fig. \ref{fig:aniso}(b), it can be observed that by changing the sample thickness from 1 cm to $0.1$ cm, the observed color after the analyzer becomes different. 
 These results illustrate that, GO locally is in a nematic phase, therefore the incident light can pass through the analyzer. 
We can induce the LC phase by applying the flow in the GO dispersion (see Fig. \ref{fig:aniso}(c)). In this case, the GO dispersion (at volume fraction of $C=0.17$Vol$\%$) is injected into the channel at a rate of $0.2$ml/h by a syringe pump. The channel is located between the crossed polarizers. We observe that the applied flow rate to GO varies the orientation of the directors and the environment of LC is anisotropic.

The flakes of GO-NLC are distributed randomly in a macro-size cell and in the vicinity of either mechanical or electric and magnetic external fields, GO-NLC reaches to order phase \cite{Lin17}. Subsequently, the incident light can pass through the analyzer. On the other hand, the GO-NLC's  flakes in a micro-channel are aligned due to anchoring force and cooperative alignment effect \cite{Batista15}.

\subsection{Experimental setup}

 Now, we turn to make micro-channel to sustain the flow of GO-NLC as an almost ideal boundary.  Such channel is prepared by soft lithography technique \cite{Duffy98,McDonald2002} and it is made from polydimethylsiloxane (PDMS). Our tasks for making the micro-channel are as follows: i) Deposition of photoresist on the hard substrate, ii) Inserting photomask on the sample and exposure to UV light, iii) Post development, iv) PDMS and cross-linker mixture is poured over the substrate with photoresist, v) The PDMS layer is thermally cured and pulled out from the substrate, vi) Producing the PDMS-glass micro-channel, creating a surface-to-wall interface between glass and PDMS, specifying inlet and outlet channels.
Finally, the thickness, width, and length of our micro-channels
are $100\pm 20 \mu$m , $300\pm 5 \mu$m and $2.0\pm  0.1$ cm, respectively. 
The channel is connected to a syringe pump and sink with an inlet and outlet tubes. 

\begin{figure}[t]
		\centering
		\includegraphics[width=10cm, height=5cm]{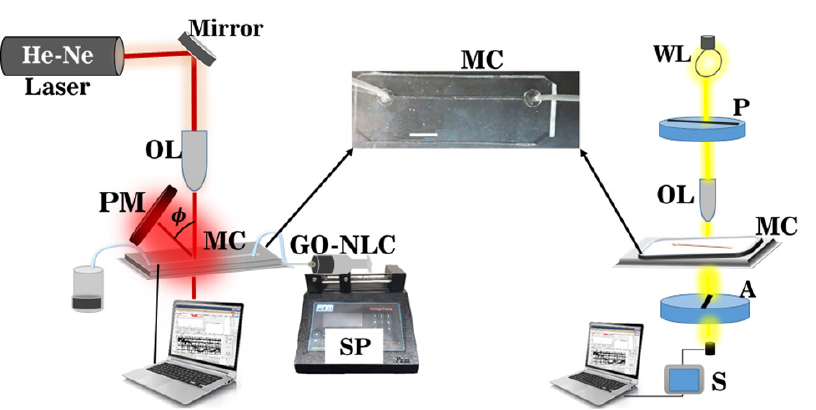}
		\caption{\label{fig:3}Schematic of the experimental setup. Left setup corresponds to the arrangement of single-mode laser light without Polaroid. The right part indicates arrangement of white light with a crossed polarizer. The inner figure is the micro-channel that used to experiment. The abbreviations used in this plot are: White Light (WL), Polarizer (P),  Objective Lens (OL), micro-channel (MC), Analyzer (A), Spectrometer (S), Power meter (PM), Syringe pump (SP).}
	\end{figure}

The effect of GO-NLC flow rate on the scattered and transmitted light are systematically investigated using the
setup illustrated in Fig. \ref{fig:3}. A continuous wave (CW) He-Ne laser ($\lambda= 633$nm ) with the Gaussian beam
profile ($\omega=1.1$mm) is focused on the sample by an objective lens. The power of laser
measured at the place of sample is $ 9 $mW. The objective lens (with $10\times$ magnification) is
used to focus the total laser beam into the channel (the channel width is $300\mu$m). Then,
using a syringe pump, the GO-NLC is injected into the channel at different flow rates
and the intensity of scattered light measured at arbitrary angle $\phi$ (the angle between the directions of the input and scattered light) by a power meter.
The detection area of the power meter (with a diameter of 1 cm) is located at a distance of 5 cm from the channel. The collection angle of scattered light is set to about $10^{\circ}$. 
We measure the scattering laser
light at different injection rates for two volume fractions of $C=0.17$Vol$\%$ and $C=0.53$Vol$\%$. In addition, the effect of injection rate on the intensity of transmitted light at different wavelength is investigated (see the right panel of Fig. \ref{fig:3}). For this purpose, the micro-channel is placed between the two crossed polarizers, and the
variation of the light intensity at different wavelength is measured. The flow direction in the channel is in parallel to the transmissive axis of the polarizer. In this measurement, we use a white light lamp with the spectral region from $200-1200$nm. To detect the light passing through the analyzer, we use a spectrometer with a $0.5$nm spectral resolution.

We have also carried out the measurements for different angles and the same behavior is observable.

\section{Results and discussion}\label{result1}

\begin{figure}[t]
		\centering
		\includegraphics[width=5.7cm, height=3cm]{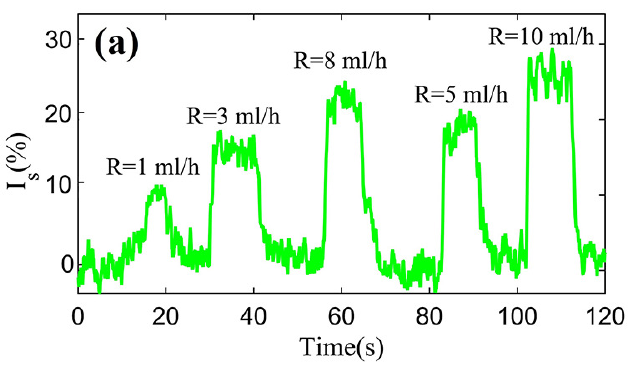}
		\includegraphics[width=5.7cm, height=3cm]{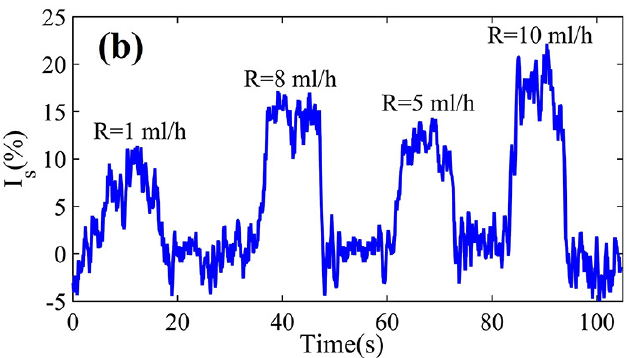}
		\includegraphics[width=5.8cm, height=3cm]{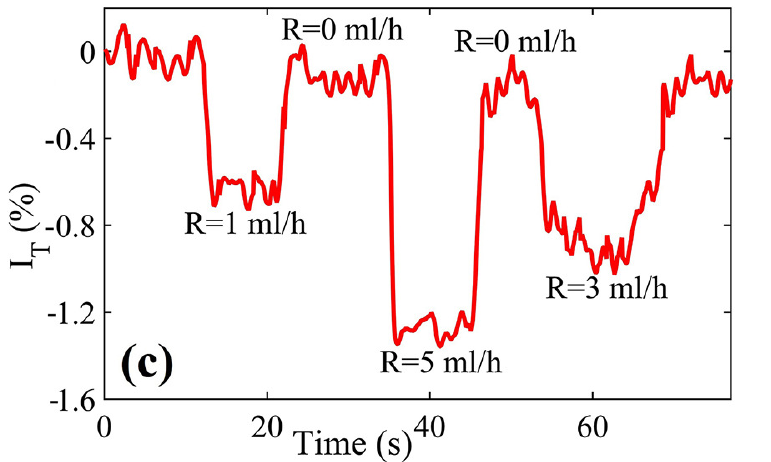}
		\includegraphics[width=5.8cm, height=3cm]{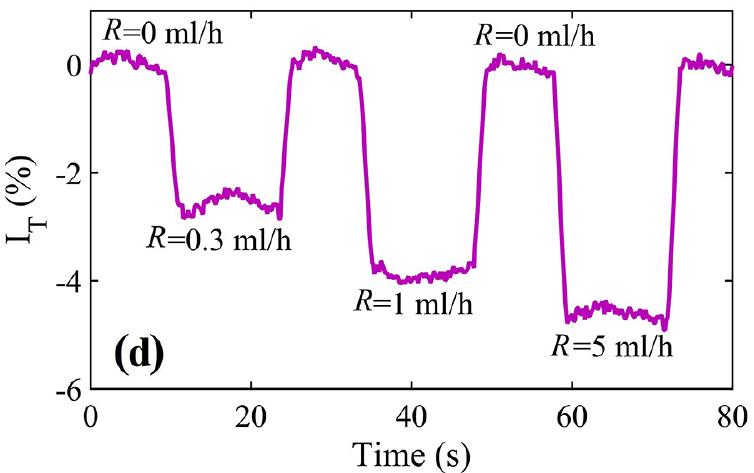}
		
		\caption{The real time measurements of the intensity of scattered light at angles of (a) $\phi= 45^{\circ}$ and (b) $\phi= 90^{\circ}$ for GO-NLC volume fractions of $C=0.17$Vol$\%$.  The real time measurement of the transmittance intensity for GO-NLC volume fractions of (c) $C=0.17$Vol$\%$ and (d) $C=0.53$Vol$\%$. }
		\label{realtime}
\end{figure}

We inject the GO-NLC with volume fractions equates to $C=0.17$Vol$\%$ and $C=0.53$Vol$\%$ into the prepared micro-channel (explained in subsection \ref{experiment1}) at different flow rates and 
we measure the normalized scattered (transmitted) laser light ($\lambda= 633$nm) intensity defined by: 
\begin{eqnarray}
\label{eq0}
I_{s,T}\equiv\frac{I_R-I_0}{I_0}\times100
\end{eqnarray}
where $I_R$ is the intensity of the scattered (transmitted) light
when the fluid is injected into the channel at a typical rate ($R$), $I_0$ is also the intensity of the scattered (transmitted) light for a situation in which the fluid has no movement (see Fig. \ref{fig:3}). Figure \ref{realtime} indicates the real time measurements of the normalized scattered light intensity for two arbitrary angles, $\phi= 45^{\circ}$ and $\phi= 90^{\circ}$, at the volume fraction of $C = 0.17$Vol$\%$. As can be observed in this figure, the normalized scattered light intensity grows by increasing the injection rate (Figs. \ref{realtime}(a) and \ref{realtime}(b)). Also, the normalized intensity of the transmitted laser light decreases by increasing the injection rate which is consistent result due to the complementary relation between scattered and transmitted parts  (Figs. \ref{realtime}(c) and \ref{realtime}(d)).

\begin{figure}[t]
		\centering
		\includegraphics[width=5.5cm, height=3.5cm]{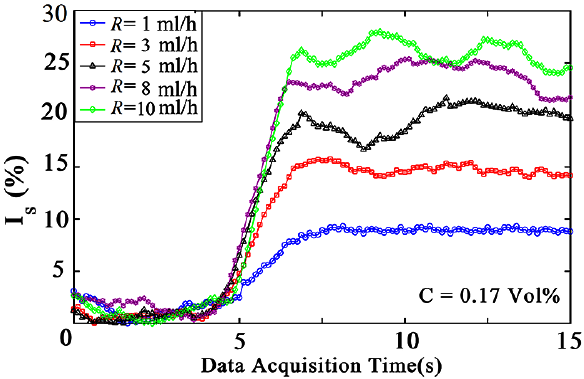}
		\includegraphics[width=5.5cm, height=3.5cm]{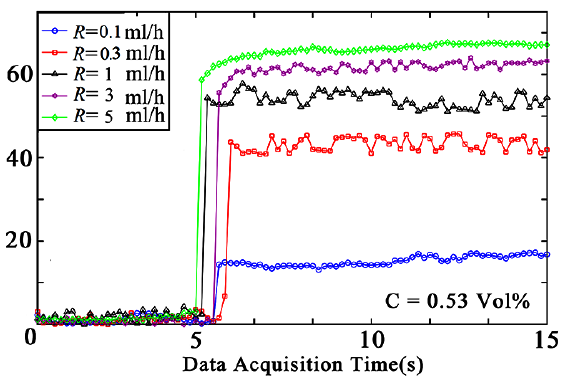}
		\caption{The intensity of scattering light for different GO-NLC injection rate into the channel for two volume fractions of $C=0.17$Vol$\%$  and $C=0.53$Vol$\%$. }
		\label{scat}
\end{figure}
\begin{figure}[t]
		\centering
		\includegraphics[width=7cm, height=4.5cm]{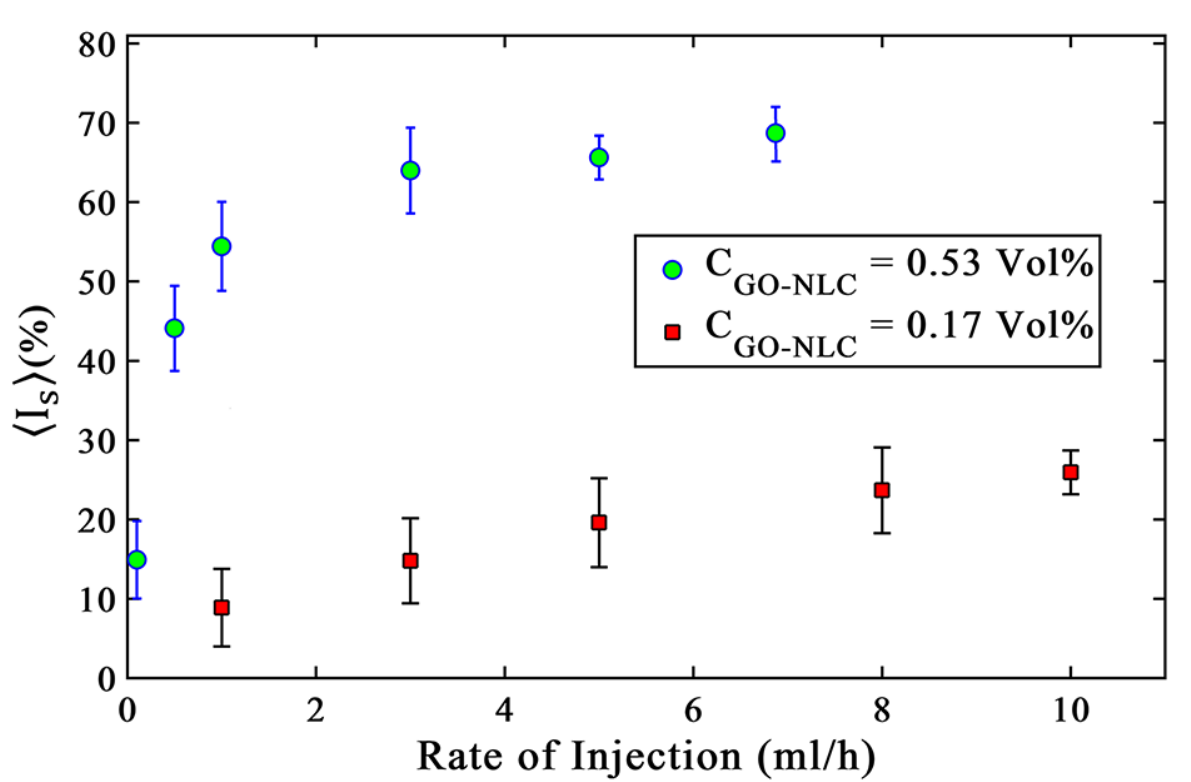}
		\caption{Experimental results for laser light scattering from GO-NLC inside a micro-channel for different injection rates at two different volume fractions of GO.}
		\label{result}
\end{figure}

It is also illustrated in Fig. \ref{scat} that the normalized scattered light intensity increases by increasing the injection rate for both volume fractions of $C=0.17$Vol$\%$ and $C=0.53$Vol$\%$. Our results indicate that for low volume fraction, the intensity reaches its saturation value slower than that of for high volume fraction. 
We also compute the time-averaged of normalized scattered light intensity, $\langle I_s\rangle$, and associated variance, $\sigma_s^2$, for different volume fractions at various injection rates. Figure \ref{result} represents $\langle I_s\rangle$ and $\sigma_s^2$ as a function of injection rate for two different volume fractions. We find that $\langle I_s\rangle$ is an increasing function versus injection rate. This phenomenon is based on the scattering of light from inhomogeneities in the dielectric constant $\left(\delta \varepsilon({\bf r},t)\right)$ of a medium that light passes through. In isotropic liquid, fluctuation of the dielectric constant is mainly due to the density variation. For liquid crystals, it should be mentioned that an additional and important factor arises from director axis variations associated with the molecular anisotropy contributes in the fluctuation of the dielectric tensor \cite{Khoo07,Yilmaz12}. The interactions in the dispersion system can be classified into flake-to-liquid and flake-to-flake interactions. When the graphene oxide volume fraction is sufficiently low, isolated flakes are the dominant form existing in the dispersion sample, and flake-to-liquid is the dominant interactions. By increasing the concentration, graphene oxide flakes start to orient due to flake-to-flake interactions. Therefore, by increasing the injection rate, the sample containing larger concentration does reorient faster and as a result, the maximum intensity of scattered light is obtained at a faster rate.

\begin{figure}[t]
		\centering
		\includegraphics[width=0.6\linewidth]{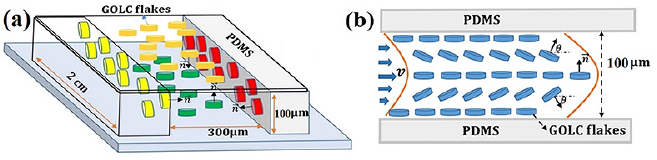}
		\includegraphics[width=0.22\linewidth]{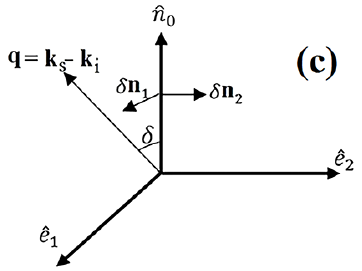}
		\caption{(a) A schematic illustration of channel characteristics and the orientations of GO-NLCs through the channel (${\bf n}$ is the director of GO's plates). (b) Poiseuille flow inside the micro-channel. (c) Coordinate system for analyzing light scattering in NLC in terms of two normal modes \cite{Khoo07}.}
		\label{flow}
\end{figure}

In order to justify the results indicated by Figs. \ref{scat} and \ref{result}, we rely on the general theoretical properties of light scattering cross-section incorporating the fluctuation of director vector and dielectric tensor. 
When the liquid is pumped through the channel, the associated directors
 are reoriented due to the external force (see Fig. \ref{flow} for more details). As shown in the mentioned 
figure, in the vicinity of the boundaries, the GO's flakes stick to the walls due to the anchoring force, and
the directors are perpendicular to the corresponding boundaries. Therefore, directional orientation behavior leads to having flakes with different alignments near and far from the  boundaries (except at the middle of the channel due to the symmetry of fluid rate) \cite{Sengupta12}.  
We suppose that the director vector can be written as: $ {\bf n}({\bf r})={\bf n}_0+\delta {\bf n}({\bf r})$. Here ${\bf n}_0$ is the initial orientation of director vector and the flow rate is different across the channel and ${\bf r}$ is the position vector, we have a fluctuation in the orientation of the directors represented by $\delta {\bf n}({\bf r})$. It turns out that, the fluctuation parallel to the initial direction of the director can not produce the relevant physical effect and therefore $\delta {\bf n}({\bf r})$ has only two components perpendicular to the ${\bf n}_0$ as shown in the right part of Fig. \ref{flow} by $\delta {n}_1$ and $\delta{n}_2$. In addition, the dielectric tensor is given by \cite{Khoo07}:  
\begin{eqnarray}
\label{eq2-10}
 {\varepsilon}_{ij} = \varepsilon_\perp \delta_{ij} + \varepsilon_a n_{i} n_{j}
\end{eqnarray}
where $\delta_{ij}$ is Kronecker function, $n_{i}$ and  $n_{j}$ are the components of director and  $\varepsilon_a \equiv\varepsilon_ \parallel -\varepsilon_\perp$. In this relation, $\varepsilon_ \parallel$ and $\varepsilon_\perp$ are the components of 
the dielectric permittivity that are parallel and perpendicular to the optical axis, respectively \cite{Blinov11}.

The fluctuations in $\varepsilon_{ij}$ ($\delta \varepsilon_{ij}({\bf r})$) come from both  the changes in $\varepsilon_\perp$ and $\varepsilon_ \parallel$ due to density variation and also due to fluctuation in director, ${\bf n}$, \cite{Khoo07}. We assume the small volume $dV$ of sample is located at the position ${\bf r}$, the induced dielectric polarization density $\delta {\bf P}$ created by applying the incident field, ${\bf E}_{in}=\hat i E_{in} \exp\left[i({\bf k}_i\cdot{\bf r}-\omega t)\right]$, reads as: $\delta {\bf P}=\varepsilon_0 \chi {\bf E}_{inc} = \delta \varepsilon {\bf E}_{inc}$. In which, $\hat{i}$ represents the direction of incident light polarization and ${\bf k}_i$ is associated  wave vector, the $\chi$ is the electric susceptibility and $\delta \varepsilon$  is the change in the dielectric constant tensor. The total outgoing component in the far field zone, ${\bf E}_{out}({\bf r}^{\prime})$, is obtained by integrating over the interaction volume $V$, as
${\bf E}_{ out}({\bf r}^{\prime})=\hat{s}  E_{out} {\exp}\left[i\left({\bf k}_s \cdot {\bf r}'-\omega t  \right )\right]$, here the $\hat{s}$ and ${\bf k}_s$ are respectively  the direction of scattered light polarization and corresponding wave vector. One can write $\hat{s} \cdot   {\bf E}_{ out}= \delta \alpha_{s} \frac{ E}{ R}  \exp\left[{i {\bf k}_s \cdot {\bf r}^{\prime}}\right]$, and $\delta \alpha_{s} $ is the perturbation part of scattered amplitude reads as \cite{Khoo07}:
\begin{eqnarray}
\label{eq2}
\delta \alpha_{s}=\frac{\omega^2}{c^2}  \int_{\rm Vol} \left[ \hat s\cdot\delta {\varepsilon}({\bf r}) \cdot\hat i^{\dagger }\right]   {\bf e}^{-i {\bf q}\cdot{\bf r}} dV 
\end{eqnarray}
where the superscript "$\dagger $" refers to the transpose vector and ${\bf q}\equiv{\bf k}_s-{\bf k}_i$. The differential scattering cross-section (the scattered power per unit incident intensity per unit solid angle) is given by \cite{Khoo07}:
\begin{eqnarray}
\label{eq6}
\frac{d\sigma}{d\Omega}\equiv\langle |\delta \alpha_s|^2\rangle =\left(\frac{\varepsilon_a \omega^2}{c^2}\right)^2   V \sum_{\alpha=1,2} \langle |\delta n_\alpha ({\bf q})|^2\rangle \left[ (\hat {n}_0\cdot\hat{i}) (\hat {e}_\alpha\cdot\hat{s})+ (\hat {n}_0\cdot\hat{s}) (\hat {e}_\alpha\cdot\hat{i})\right]^2
\end{eqnarray}
where $\sigma$ is the scattering cross-section (the scattered power per unit incident intensity at scattering angle) and $\Omega$ is the solid angle. In this equation, the director fluctuation $\langle | \delta n_{\alpha}({\bf q})|^2\rangle$ can be obtain by averaging in Fourier space as follows \cite{Marusii86}:
\begin{eqnarray}
\label{eq5}
\langle | \delta n_{\alpha}({\bf q})|^2\rangle =\frac{1}{V}\frac{k_BT}{K_{\alpha} q_1^2+K_3q_3^2}
\end{eqnarray}
where $k_B$ is the Boltzman's constant, $T$ is the temperature of the NLC and $V$ is the volume of the investigated area. The volume should be taken into account is the interaction volume. Therefore, for our experimental setup, it can be considered as a cylindrical volume with the height corresponds to the channel depth and its cross-section is equal to the cross-section of the laser beam at the interaction zone. Also, $K_{1}$, $K_{2}$ and $K_{3}$ are elastic constants of Splay, Twist, and Bend affecting on the  NLC director and they change the ${\bf n}$ due to the flow rate.

According to Eq. (\ref{eq6}), the cross-section of scattering can be changed due to fluctuation of the directors, $\langle |\delta n_\alpha ({\bf q})|^2\rangle$, and  the dielectric matrix component $\varepsilon_a= n_e ^ 2-n_o ^ 2$ depending on NLC birefringence \cite{Sengupta12-1}. Both mentioned components are affected by rate of flow injection. We expect to have an increasing behavior in fluctuation of the director by increasing the injection rate. As we will demonstrate in the following part, the amount of birefringence also increases by increasing $R$. Subsequently, our experimental results can be justified by theoretical achievement (Figs. \ref{realtime}-\ref{result}).  

To check the dependency of the birefringence on the injection rate, we experimentally evaluate  the white light transmission from the micro-channel containing GO-NLC. To this end, we place the micro-channel between the crossed polarizers and the un-polarized light of halogen lamp (with the spectrum of $200$nm to $1200$nm) focused on the channel as illustrated in the right part of Fig. \ref{fig:3}.

\begin{figure}[t]
		\centering
		\includegraphics[width=7cm, height=5cm]{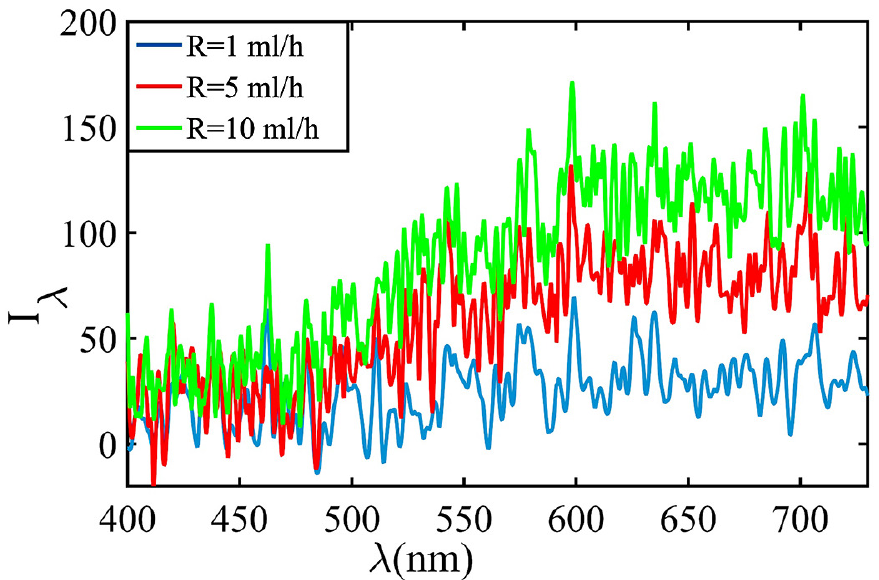}
		\caption{ The output intensity of the transmitted light from a micro-channel containing GO-NLC for injection rates of 1ml/h, 5ml/h and 10ml/h, where the channel is placed between crossed polarizer.}
		\label{intensity}
\end{figure}

The Fig. \ref{intensity} shows the normalized transmitted intensity in the visible region for three injection
rates namely, 1ml/h, 5ml/h, and 10ml/h. The normalized transmitted light intensity at wavelength of $\lambda$  is defined as:
\begin{eqnarray}
\label{eq12}
I_\lambda \equiv\frac{I_{R,\lambda}-I_{0,\lambda}}{I_{0,\lambda}}\times100
\end{eqnarray}
where $I_{R,\lambda}$  and $I_{0,\lambda}$  are the intensity of the transmitted light with and without flow rate, respectively.
One can observe that by increasing the injection rate of GO-NLC into the
channel, the intensity of the beam passing through the analyzer almost increases in the interval of spectral range $[400-800]$nm. According to the lower panel of Fig. \ref{realtime}, the higher flow rate causes to obtain the lower transmittance intensity before analyzer and simultaneously, an increasing in the scattered intensity is detected. In the absence of crossed polarizer, a decreasing behavior in the transmitted intensity is recognized when the scattered intensity grows. In other word, under no injection rate, the GO-NLC sample is isotropic, consequently, we observe the dark state under crossed polarizer. Applying the injection rate, the GO-NLC flakes are partially aligned, and the transmission would be increased due to the induced birefringence in the samples. Generally, the GO-NLC cell alters the linear polarized light into elliptically polarized light. The state of elliptical polarization depends on the sample thickness, wavelength and birefringence. Therefore, by increasing the birefringence, polarization gets higher variations. As a result, the transmission of the light from the analyzer is increased which is equivalent to the lower transmission of light when we didn't use analyzer (see parts "c "and "d " of Fig. \ref{realtime}). During the data acquisition in the experiment, there are many reasons to obtain shot noises and outliers sourced by systematic effects. To reduce the impact of mentioned events, we have used a kernel function and finally we reported convolved data. Taking into account more realization for ensemble averaging, may suppress and alleviate most parts of artificial noises and fluctuations, nevertheless, examining the details of variations at different wavelengths needs to more precise experimental setup.

This phenomenon can be explained as follows: the flow of GO-NLC changes the direction of the director leading to increase the birefringence. The birefringence growth with respect to the flow rate reaches a saturation amount. Therefore, one can write \cite{Sengupta12-1}:
\begin{eqnarray}
\label{eq8}
n_o=n_\parallel &{\rm and}&   n_e=\frac{n_\parallel n_\perp}{\sqrt{n_\parallel ^2 \cos^2(\theta)+n_\perp ^2 \sin^2(\theta)}}
\end{eqnarray}
where $n_\parallel$ and $n_\perp$ are the refractive index parallel  and perpendicular to the director, respectively. The $\theta$ shows the angle between the wave vector of the incident light and the optical axis. By increasing the injection rate of GO-NLC into the channel, $\theta$ deviates from its initial value.  
Consequently, according to Eq. (\ref{eq8}), the difference between $n_o$ and $n_e$ is enhanced in different regions of the channel. Therefore, the variation in the polarization of the light passing through the medium essentially changes the intensity of transmitted light. 
The variation in the director angle deviation has a maximum value (saturation angle) which is proportional to the hydrodynamic coefficient of the GO-NLC:
\begin{eqnarray}
\label{eq81}
\cos(2\theta_{s})=-\frac{\gamma_2-\gamma_3}{\gamma_5-\gamma_6}
\end{eqnarray}
 where $\theta_{s}$ is the saturation angle and $\gamma_i$s are the Leslie viscosity coefficients for anisotropic fluid \cite{Blinov11, Leslie85, Leslie68}. Therefore, the variation of scattering cross-section as a function of injection
rate must be saturated (see Fig. \ref{result} to make more sense).

\section{Conclusion}\label{summary1}

The interaction of light passing through a liquid crystal in the nematic phase can be a probe to examine the optical properties which is utilized for controllability behavior. Mechanical-hydrodynamical approaches including flow shear to manipulate the orientation of graphene-based materials have no enough flexibility, and therefore alternative techniques such as applying the electric or magnetic fields on the underlying NLC have specific impact \cite{Song16,Ahmad15,Ahmad16,Hong15,Nehring76,Ong84,Brochard72,Doorn75}. Nevertheless, the  low-cost and feasible as well as simple experimental setup reached by mechanical-hydrodynamical methods motivated us to concentrate on the mentioned approach and we tried to improve its imperfections such as amount of controllability in orientation of associated flakes of GO-NLC as a case study.  Accordingly, we could make a tunable functional material based on liquid crystal in the nematic phase by adopting the flow rate in a micro-channel.   

We injected the GO-NLC into the micro-channel made in our laboratory by a syringe pump with various flow rates at volume fractions equates to $C=0.17$Vol$\%$ and $C=0.53$Vol$\%$ (see Fig. \ref{fig:3}). Meanwhile, we measured the normalized scattered and transmitted light intensities by a power meter as a function of time (see Fig. \ref{realtime}). The time averaged of scattered light intensity fluctuations was an increasing function with respect to volume fraction. The intensity of scattered light reaches its saturation values for higher volume fraction sooner than that of for lower volume fraction.       

Utilizing the theoretical model for cross-section of scattering amplitude by means of mathematical description of fluctuations in director vector and dielectric tensors enabled us to explain the behavior of scattered light intensity with respect to the flow rate achieving a tunable matter by mechanical-hydrodynamical force. Variation in the fluid rate through the micro-channel had an impact on the director fluctuations and the dielectric matrix. According to our setup for examining the transmitted light, we found that birefringence property depends on the injection rate, directly. The growing in birefringence reached a saturation value. By increasing the injection flow rate, the angle between the wave vector of incident light ray and optical axis got more variation with respect to its initial situation leading to an enhancement in absolute difference between $n_o$ and $n_e$ consequently, the polarization of light passing through the media varied. Finally, we observed an increase in intensity of transmitted light  passing through the sample that placed between crossed polarizer (Fig. \ref{intensity}). We point out that the mentioned result  is consistent with simultaneous increasing behavior for scattered intensity since we have utilized analyzer. In the absence of crossed polarizer, a decreasing behavior in transmitted intensity is recognized when the scattered intensity grows (Fig. \ref{realtime}).

We summarize that in this paper, we constructed an experimental setup to achieve a tunable functional material based on NLC phase utilizing mechanical-hydrodynamical force in a micro-channel. To the best of our knowledge, our approach is one of the feasible and robust setups taking into account mechanical method rather than applying the electric or magnetic field to achieve an acceptable degree of controllability. Considering the impact of additional physical quantities such as variation of temperature, the shape of micro-channel and viscosity of sample and size of flakes on the interaction of incident light ray with NLC phase are interesting and can be examined in the future work.

\section*{Funding}

Deputy for Research and Technology of Shahid Beheshti University (600/5290).
 %{\bf Acknowledgement:}
 %The authors thanks to anonymous referee %for his/her useful comments which enabled %us to improve this paper.   
 
%\bibliography{references}

\bibliography{Reference}

\end{document}